\documentstyle[preprint,prd,aps,eqsecnum,amssymb]{revtex}

\tighten

\begin{document}
\preprint{DAMTP-R94/35}
\title{Unitarity of Quantum Theory and \\ Closed Time-like Curves}
\author{C.J. Fewster\thanks{E-mail address: fewster@butp.unibe.ch}}
\address{Department of Applied Mathematics and Theoretical Physics,
University of Cambridge, \\Silver Street, Cambridge CB3 9EW, U.K.
\\ and Institut f\"{u}r theoretische Physik, Universit\"{a}t Bern, \\
Sidlerstrasse 5, CH-3012 BERN, Switzerland.\footnote{Current Address}}
\author{C.G. Wells\thanks{E-mail address: C.G.Wells@damtp.cam.ac.uk}}
\address{Department of Applied Mathematics and Theoretical Physics,
University of Cambridge, \\ Silver Street, Cambridge CB3 9EW, U.K.}
\date{15 November 1994, Revised \today}
\maketitle

\begin{abstract}

Interacting quantum fields on spacetimes containing regions of closed
timelike curves (CTCs) are subject to a non-unitary evolution $X$.
Recently, a prescription has been proposed, which restores unitarity
of the evolution by modifying the inner product on the final Hilbert
space. We give a rigorous description of this proposal and note an
operational problem which arises when one considers the composition
of two or more non-unitary evolutions.

We propose an alternative method by which unitarity of the
evolution may be regained, by extending $X$ to a unitary evolution on
a larger (possibly indefinite) inner product space. The proposal
removes the ambiguity noted by Jacobson in assigning expectation
values to observables localised in regions spacelike separated from
the CTC region. We comment on the physical significance of the
possible indefiniteness of the inner product introduced in our
proposal.

\end{abstract}


\narrowtext

\section{Introduction}

Various recent studies \cite{Boul,FPS1,Pol} of perturbative
interacting quantum field theory in the presence of a compact region
of closed timelike curves (CTCs) have concluded that the evolution
from initial states in the far past of the CTCs to final states in
their far future fails to be unitary, in contrast with the situation
for free fields \cite{Boul,FPS2,GPPT}. The same conclusion has also
been reached non-perturbatively for a model quantum field theory
\cite{Pol2}. This presents many problems for the usual Hilbert space
framework of quantum theory: as we describe in
Section~\ref{sect:nuqm}, the Schr\"{o}dinger and Heisenberg pictures
are inequivalent and ambiguities arise in assigning probabilities to
events occurring before \cite{FPS1}, or spacelike separated from
\cite{Jacobson}, the region of non-unitary evolution.

The main reaction to these difficulties has been to abandon the
Hilbert space formulation in favour of a sum over histories approach
such as the generalised quantum mechanics of Gell-Mann and Hartle
(see, e.g., \cite{GMH}).  In particular, Hartle \cite{Hartle} has
addressed the issue of non-unitary evolutions in generalised quantum
mechanics.  Nonetheless, it is of interest to see if the Hilbert
space approach can be `repaired' by restoring unitarity.  Recently,
Anderson \cite{Arley} has proposed that this be done as follows.
Suppose a non-unitary evolution operator $X$ is defined on Hilbert
space ${\cal H}$ with inner product $\langle\cdot\mid\cdot\rangle$.
We assume that $X$ is bounded with bounded inverse. Anderson defines
a new inner product $\langle\cdot\mid\cdot\rangle^\prime$ on ${\cal
H}$ by $\langle\psi\mid\varphi\rangle^\prime = \langle X^{-1}\psi \mid
X^{-1}\varphi \rangle$, and denotes ${\cal H}$ equipped with the new
inner product as ${\cal H}^\prime$. Regarded as a map from ${\cal H}$
to ${\cal H}^\prime$, $X$ is clearly unitary.\footnote{We will give a
more rigorous formulation of this statement in
Section~\ref{sect:And}.} The essence of Anderson's proposal is to
restore unitarity by regarding $X$ in this way. Of course, one also
needs to be able to represent  observables as self-adjoint operators
on both Hilbert spaces; Anderson has shown how this may be done by
establishing a correspondence (depending on the evolution) between
self-adjoint operators on ${\cal H}$ and those on ${\cal H}^\prime$.
When only one non-unitary evolution is considered, this proposal is
equivalent to remaining in the Hilbert space ${\cal H}$ and replacing
$X$ by $U_X = (XX^*)^{-1/2}X$, i.e., the unitary part of $X$ in the
sense of the polar decomposition~\cite{RSi}.

A curious feature of Anderson's proposal emerges when one considers
the composition of two or more consecutive periods of non-unitary
evolution~\cite{Anote}. If an evolution $Y$ is followed by $X$,
one might expect that the combined evolution would be
represented by the composition of the unitary parts, i.e., $U_X
U_Y$. However, this does not generally agree with the unitary
part of the composition, $U_{XY}$, and so there would be an
ambiguity depending on whether one thought of the full evolution
as a one-stage or two-stage journey. Anderson's response to this
is to argue that the second evolution should be treated in a
different way, essentially (as we show in Section~\ref{sect:And})
by replacing $X$ by the unitary part of $X(YY^*)^{1/2}$. This
removes the ambiguity mentioned above, but has the undesirable
feature that the treatment of the second evolution depends on
the first. In Section~\ref{sect:And}, we will show that this
leads to an operational problem for physicists living in a
universe containing CTC regions.

It is therefore prudent to seek other means by which unitarity can be
restored. In this paper, we propose a method of
unitarity restoration using the mathematical technique of {\em
unitary dilations}. This is motivated by the simple geometric
observation that any linear transformation of the real line is the
projection of an orthogonal transformation (called an {\em orthogonal
dilation} of the original mapping) in a larger (possibly indefinite)
inner product space. To see this, note that any linear contraction on
the line may be regarded as the projection of a rotation in the
plane: the contraction in length along the $x$-axis, say, being
balanced by a growth in the $y$-component. Similarly, a linear
dilation on the line may be regarded as the projection of a Lorentz
boost in two dimensional Minkowski space. This observation may be
extended to operators on Hilbert spaces: it was shown by Sz.-Nagy
\cite{Nagy} that any contraction (i.e., an operator $X$ such that
$\|X\psi\|\le\|\psi\|$ for all $\psi$) has a unitary dilation acting
on a larger Hilbert space. The theory was subsequently extended to
non-contractive operators by Davis \cite{Davis} at the cost of
introducing indefinite inner product spaces. Unitary dilations have
previously found physical applications in the quantum theory of open
systems \cite{Davies}, and have also been employed by one of us in an
inverse scattering construction of point-like interactions in quantum
mechanics \cite{F1,F2}.

Put concisely, starting with a non-unitary evolution $X$, we pass to
a unitary dilation of $X$, mapping between enlarged inner product
spaces whose inner product may (possibly generically) be
indefinite. The signature of the inner product is determined by the
operator norm $\|X\|$ of $X$: if $\|X\|\le 1$, the enlarged inner
product spaces are Hilbert spaces, whilst for $\|X\|>1$, they are
indefinite inner product spaces (Krein spaces). Within the context of
our proposal, it is therefore important to determine $\|X\|$ for any
given CTC evolution operator.

Essentially, the unitary dilation proposal performs the minimal
book-keeping required to restore unitarity by asserting the presence
of a hidden component of the wavefunction, which is naturally
associated with the CTC region. These `extra dimensions' are not
accessible to experiments conducted outside the CTC region, but
provide somewhere for particles to hide from view, whilst
maintaining global unitarity. We will see that our proposal thereby
circumvents the problems associated with non-unitary
evolutions mentioned above.

Of course, it is a moot point whether or not one should require a
unitary evolution of quantum fields in the presence of CTCs; one
might prefer a more radical approach such as that advocated by Hartle
\cite{Hartle}. Our philosophy here is to determine the extent to
which the conventional formalism of quantum theory can be repaired.

The plan of the paper is as follows. We begin in
Section~\ref{sect:nuqm} by describing the implications of
non-unitarity for the Hilbert space formulation of quantum mechanics
and then give a rigorous description of Anderson's proposal in
Section~\ref{sect:And}, where we also note the operational problem
mentioned above. In Section~\ref{sect:udp}, we
introduce our proposal for unitarity restoration, and show how
composition may be treated within this context in
Section~\ref{sect:comp}. In Section~\ref{sect:conc}, we conclude
by discussing the physical significance of our proposal. There
are two appendices: Appendix~A contains the proof of two results
required in the text, whilst Appendix~B describes yet another
proposal for unitarity restoration based on tensor products. However,
this proposal (in contrast to that advocated by Anderson, and
our own) fails to remove the ambiguity noted by
Jacobson~\cite{Jacobson}.

\section{Non-Unitary Quantum Mechanics}
\label{sect:nuqm}

As we mentioned above, a non-unitary evolution raises many problems
for the standard formalism and interpretation of quantum theory,
some of which we now discuss.

Firstly, the usual equivalence of the Schr\"{o}dinger and Heisenberg
pictures is lost. Given an evolution $X$ of states and an
observable $A$, we would naturally define the evolved observable
$A^\prime$ so that for all initial states $\psi$, the expectation
value of $A^\prime$ in state $\psi$ equals the expectation of $A$ in
the evolved state $X\psi$. Explicitly, we require
\begin{equation}
\frac{\langle{\psi}\mid{A^\prime\psi}\rangle}{\langle\psi\mid\psi
\rangle}=
\frac{\langle X\psi\mid AX\psi\rangle}{\langle X\psi\mid X\psi
\rangle}
\label{eq:exp}
\end{equation}
for all $\psi$ in the Hilbert space ${\cal H}$. If $X$ is unitary up
to a scale (i.e., $X^*X=XX^*=\lambda\openone$, $\lambda\in{\Bbb
R}^+$), then equation~(\ref{eq:exp}) is uniquely solved by the
Heisenberg evolution $A^\prime=X^{-1}AX$. On the other hand, if $X$
is not unitary up to scale, then there is no operator $A^\prime$
satisfying~(\ref{eq:exp}) unless $A$ is a scalar multiple of the
identity.

For completeness, we give a proof of this fact. Defining $f(\psi)$ to
equal the RHS of~(\ref{eq:exp}), and taking $\psi$ and $\varphi$ to
be any orthonormal vectors, we note that linearity of $A^\prime$
entails
\begin{equation}
f(\psi)+f(\varphi)=f(\psi+\varphi)+f(\psi-\varphi),
\end{equation}
whilst linearity of $A$ implies
\begin{eqnarray}
f(\psi)\|X\psi\|^2 + f(\varphi)\|X\varphi\|^2
&=&\frac{1}{2}\left\{f(\psi+\varphi)\|X(\psi+\varphi)\|^2 \right.
\nonumber \\
&&\!\!\!\!\!\!\!\!\!\!\!\!\!\!\!\!+\left.
f(\psi-\varphi)\|X(\psi-\varphi)\|^2\right\}.
\end{eqnarray}
Multiplying $\varphi$ by a phase to ensure that $\langle X\psi\mid
X\varphi\rangle$ is imaginary (and hence that
$\|X(\psi\pm\varphi)\|^2=\|X\psi\|^2+\|X\varphi\|^2$), we combine
these relations to obtain
\begin{equation}
(f(\psi)-f(\varphi))(\|X\psi\|^2-\|X\varphi\|^2)=0 ,
\end{equation}
which is clearly insensitive to the phase of $\varphi$ and therefore
holds for all orthonormal vectors $\psi$ and $\varphi$. If $X$ is not
unitary up to scale, we choose $\varphi$ and $\psi$ so that
$\|X\psi\|\not=\|X\varphi\|$. Thus $f(\psi)=f(\varphi)=F$ for some
$F$. It follows that $f(\chi)=F$ for all $\chi\perp{\rm
span}\,\{\psi,\varphi\}$ (because $\|X\chi\|$ cannot equal both
$\|X\psi\|$ and $\|X\varphi\|$) and hence for all $\chi\in{\cal H}$.
Thus $A$ is a scalar multiple of the identity.

Thus, the conventional equivalence of the Schr\"{o}dinger and
Heisenberg pictures is radically broken. If there are evolved states,
there are no evolved operators, and {\em vice versa}. In addition,
the Heisenberg picture places restrictions on the class of allowed
observables. In order to preserve the canonical commutation
relations, we take the evolution to be $A\rightarrow X^{-1}AX$;
however, we also want to preserve self-adjointness of observables
under evolution. Combining these two requirements, we conclude that
$A$ must commute with $XX^*$ and therefore with $(XX^*)^{1/2}$ -- the
non-unitary part of the evolution in the sense of the polar
decomposition. Thus, the claim attributed to Dirac \cite{Rovelli}
that `Heisenberg mechanics is the good mechanics' carries the price
of a restricted class of observables when the evolution is
non-unitary.

A second problem with non-unitary evolutions, noted by Jacobson
\cite{Jacobson} (see also Hartle's elaboration \cite{Hartle}) is that
one cannot assign unambiguous values to expectation values of
operators localised in regions spacelike separated from the CTC
region. Let $\cal R$ be a compact region spacelike separated from the
CTCs, and which is contained in two spacelike slices $\sigma_+$ and
$\sigma_-$, such that $\sigma_-$ passes to the past of the CTCs and
$\sigma_+$ to their future. If $A$ is an observable which is
localised within $\cal R$, one can measure its expectation value with
respect to the wavefunction on either spacelike surface. In order for
these values to agree, equation~(\ref{eq:exp}) must hold with
$A^\prime=A$. If $X$ is unitary up to scale, this is satisfied by any
observable which commutes with $X$ -- in particular by all
observables localised in $\cal R$. However, if $X$ is not unitary up
to scale, our arguments above show that there is no observable (other
than multiples of the identity) for which unambiguous expectation
values may be calculated. Jacobson concludes that a breakdown of
unitarity implies a breakdown of causality.

Thirdly, Friedman, Papastamatiou and Simon \cite{FPS1} have pointed
out related problems with the assignment of probabilities for events
occurring before the region of CTCs. They consider a microscopic
system which interacts momentarily with a measuring device before the
CTC region and which is decoupled from it thereafter. The microscopic
system passes through the CTC region, whilst the measuring device
does not. However, the probability that a certain outcome is
observed on the measuring device depends on whether it is observed
before or after the microscopic system passes through the CTCs.
This is at variance with the Copenhagen interpretation of quantum
theory.

\section{The Anderson Proposal}
\label{sect:And}

We begin by giving a rigorous description of Anderson's
proposal~\cite{Arley}. Let
${\cal H}$ be a Hilbert space with inner product $\langle\cdot
\mid\cdot \rangle$ and suppose that the non-unitary evolution operator
$X:{\cal H} \rightarrow {\cal H}$ is bounded with bounded inverse.  We
now define a quadratic form on ${\cal H}$ by
\begin{equation}
q(\psi,\varphi) = \langle X^{-1}\psi
\mid X^{-1}\varphi \rangle ,
\end{equation}
which (because $(X^{-1})^*X^{-1}$ is positive and $X$ and $X^{-1}$
are bounded) defines a positive definite inner product on ${\cal H}$
whose associated norm is complete. Replacing
$\langle\cdot\mid\cdot\rangle$ by this inner product, we obtain a new
Hilbert space which we denote by ${\cal H}^\prime$. Because ${\cal
H}^\prime$ coincides with ${\cal H}$ as a vector space, there is an
identification mapping $\i:{\cal H}
\rightarrow{\cal H}^\prime$ which maps $\psi\in{\cal H}$ to
$\psi\in{\cal H}^\prime$. The inner product of ${\cal H}^\prime$ is
\begin{equation}
\langle\psi\mid\varphi\rangle^\prime = \langle X^{-1}\i^{-1}\psi
\mid X^{-1}\i^{-1}\varphi \rangle , \label{eq:IPprime}
\end{equation}
for $\psi,\varphi\in{\cal H}^\prime$. The identification mapping is
present because $X^{-1}$ is not, strictly speaking, defined on ${\cal
H}^\prime$. As a minor abuse of notation, one can omit these
mappings provided that one takes care of which inner product and
adjoint are used in any manipulations. This is the approach
adopted by Anderson. The advantage of writing in the identifications
is that one cannot lose track of the domain or range of any operator,
and adjoints automatically take care of themselves.

{}From equation~(\ref{eq:IPprime}), it is clear that $\i X:{\cal
H}\rightarrow {\cal H}^\prime$ (i.e., ``$X$ regarded as a map from
${\cal H}$ to ${\cal H}^\prime$'') is unitary -- the non-unitarity of
$X$ is cancelled by that of $\i$. Anderson therefore adopts $\i X$ as
the correct unitary evolution: in the Schr\"{o}dinger picture, an
initial state $\psi\in{\cal H}$ is evolved unitarily to $\i X\psi\in
{\cal H}^\prime$.

The next component in Anderson's proposal concerns observables.
Given an observable (e.g., momentum or position) represented as a
self-adjoint operator $A$ on ${\cal H}$, one needs to know how this
observable is represented on ${\cal H}^\prime$ in order to evolve
expectation values in the Schr\"{o}dinger picture. At first, one
might imagine that $A$ should be carried over directly using the
identification mapping to form $A^\prime = \i A\i^{-1}$.  However,
this idea fails because $\i A\i^{-1}$ is not self-adjoint in ${\cal
H}^\prime$ unless $A$ commutes with $XX^*$: an unacceptable
restriction on the class of observables. Instead, Anderson proposes
that $A^\prime$ should be defined by
\begin{equation}
A^\prime = \i R_X A R_X^{-1} \i^{-1}
\label{eq:otra}
\end{equation}
where $R_X = (XX^*)^{1/2}$ is self-adjoint and positive on ${\cal
H}$.  The operator $\i R_X$ is easily seen to be unitary, and it
follows that $A^\prime$ is self-adjoint on ${\cal H}^\prime$. With
this definition, the expectation value of $A$ in (normalised) state
$\psi$ evolves as
\begin{equation}
\langle\psi\mid A\psi\rangle \longrightarrow
\langle \i X\psi \mid A^\prime \i X\psi \rangle^\prime =
\langle U_X\psi\mid AU_X\psi\rangle ,
\end{equation}
where $U_X=R_X^{-1}X$ is the unitary part of $X$ in the sense of the
polar decomposition~\cite{RSi}.

So far, it appears that Anderson's proposal is equivalent to
Schr\"{o}dinger picture evolution using $U_X$ in the original Hilbert
space, or Heisenberg evolution $A\rightarrow U_X^{-1}AU_X$. However,
one must be careful with this statement when one considers the
composition of two consecutive periods of evolution, say $Y$ followed
by $X$. We take both operators to be maps of ${\cal H}$ to itself, as
required by Anderson~\cite{Anote,Aques}. Proceeding na\"{\i}vely,
we encounter the following problem:  taking the unitary parts and
composing, we obtain $U_XU_Y$, whilst composing and taking the unitary
part (i.e., considering the evolution as a whole, rather than as a two
stage process) we find $U_{XY}$. For consistency, we would require
that these evolutions should be equal up to a complex phase $\lambda$.
As we show in Appendix~A, this is possible if and only if $X^*X$
commutes with $YY^*$ and $\lambda = 1$. Composition would therefore
fail in general.

In response to this, Anderson has proposed that composition be treated
as follows~\cite{Anote}. Suppose $Y:{\cal H}\rightarrow {\cal
H}$ is the first non-unitary evolution, and apply Anderson's proposal
to form a Hilbert space ${\cal H}^\prime$ and an identification map
$\j:{\cal H}\rightarrow {\cal H}^\prime$ so that $\j Y$ is unitary.
The next step is to form the `push-forward' $X^\prime$ of the operator
$XR_Y$ to ${\cal H}^\prime$, which is defined by
\begin{equation}
X^\prime = \j R_Y (XR_Y) R_Y^{-1}\j^{-1} = \j R_Y X \j^{-1}.
\end{equation}
$X^\prime$ is decomposed as $R_{X^\prime} U_{X^\prime}$ in
${\cal H}^\prime$,
and $U_{X^\prime}$ is `pulled back' to ${\cal H}$ as
$\widetilde{U_{X^\prime}}
 = R_Y^{-1}\j^{-1}U_{X^\prime}\j R_Y$. Anderson states that the
correct
composition law is to form the product $\widetilde{U_{X^\prime}}U_Y$.
In fact, we can simplify this slightly, because
\begin{equation}
\widetilde{U_{X^\prime}} = R_Y^{-1} \j^{-1} U_{X^\prime}\j R_Y
 = U_{R_Y^{-1}\j^{-1}X^\prime\j R_Y} = U_{X R_Y}
\end{equation}
where we have used the fact that $U_{VXW} = V U_X W$ if $V$ and $W$
are unitary. Thus we can eliminate ${\cal H}^\prime$ from the
discussion, and the composition rule is essentially to replace the
second evolution by $U_{X R_Y}$ rather than $U_X$. This is certainly
consistent: for $U_{X R_Y} = U_{XY U_Y^{-1}}= U_{XY} U_Y^{-1}$, and
so $U_{X R_Y} U_Y = U_{XY}$.

However, although this prescription is consistent, it has the drawback
that one must know about the first non-unitary evolution in order
to treat the second correctly (i.e., one must use $U_{X R_Y}$ rather
than $U_X$). More generally, it is easy to see that, given $n$
consecutive evolutions $X_1,\ldots,X_n$, one should replace each
$X_r$ by $U_{X_r R_{X_{r-1}\ldots X_2 X_1}}$ for $r\ge 1$,
so one needs to know about all previous evolutions at each step.

This gives rise to the following operational problem:
suppose two physicists, $A$ and $B$ live in a universe with two
isolated compact CTC regions corresponding to evolutions $Y$ and
$X$ respectively. Suppose that $A$ knows about both evolutions, but
$B$ only knows about $X$. Thus, if $A$ follows Anderson's proposal,
she replaces these evolutions by $U_Y$ and $U_{XR_Y}$ respectively.
But $B$ would surely replace $X$ by $U_X$, which differs from
$U_{XR_Y}$
unless $X^*X$ commutes with $YY^*$ (as a corollary of the Theorem in
Appendix~A). {\em The two physicists treat the second evolution in
different ways and will therefore compute different values for
expectation values of physical observables in the final
state.}\footnote{One {\em can} arrange that $A$ and $B$
agree if $A$ replaces $Y$ and $X$ by $U_{R_X Y}$ and $U_X$
respectively, because $U_{R_X Y} = U_X^{-1}U_{XY}$. However, this
would require $A$
to know about $B$'s existence and ignorance of the first evolution.}
This shows that, in  Anderson's proposal, it is necessary to
know about all
non-unitary evolutions in one's past in order to treat non-unitary
evolutions in one's future correctly.

For completeness, let us see how this composition law appears in the
formulation of Anderson's proposal in which one modifies the Hilbert
space inner product. Again we start with the evolution
$Y$, and form the identification map $\j:{\cal H}\rightarrow
{\cal H}^\prime$. In addition, we can treat the combined evolution
$Z =XY$ using Anderson's proposal to form a Hilbert space
${\cal H}^{\prime\prime}$ and identification
map ${\rm k}:{\cal H}\rightarrow {\cal H}^{\prime\prime}$, such that
${\rm k} Z$ is unitary. The wavefunction is evolved from ${\cal H}$ to
${\cal H}^\prime$ using $\j Y$, and from ${\cal H}$ to
${\cal H}^{\prime\prime}$ using ${\rm k} Z$. Thus it evolves from
${\cal H}^\prime$ to ${\cal H}^{\prime\prime}$ under
${\rm k}Z(\j Y)^{-1} = \i \j X\j^{-1}$, where $\i = {\rm k}\j^{-1}$ is
clearly the identification mapping between ${\cal H}^{\prime}$ and
${\cal H}^{\prime\prime}$. This evolution, which is forced upon
us by the requirement that the wavefunction be evolved consistently,
is exactly what arises from Anderson's proposal applied to the
operator $\j X\j^{-1}$ in ${\cal H}^\prime$. One might expect that
observables would be transformed from ${\cal H}^\prime$ to
${\cal H}^{\prime\prime}$ using the rule~(\ref{eq:otra}) applied
to this evolution. However, we will now show that this is not
the case.

An observable $A$ on ${\cal H}$ is represented as $A^\prime =
\j R_Y A R_Y^{-1}\j^{-1}$ on ${\cal H}^\prime$, and by
$A^{\prime\prime}=
{\rm k} R_Z A R_Z^{-1}{\rm k}^{-1}$ on ${\cal H}^{\prime\prime}$.
Thus, the transformation between $A^\prime$ and
$A^{\prime\prime}$ is
\begin{equation}
A^{\prime\prime} = {\rm k} R_Z R_Y^{-1}\j^{-1} A^\prime
\j R_Y R_Z^{-1}{\rm k}^{-1}.
\label{eq:A12}
\end{equation}
Let us note that this is {\em not} the transformation law which
follows from a na\"{\i}ve application of Anderson's proposal to
$\j X \j^{-1}$ in ${\cal H}^{\prime}$, which would be of form
\begin{equation}
A^{\prime\prime} = \i R_W A^\prime R_W^{-1}\i^{-1}
\label{eq:A12fm}
\end{equation}
with $W = \j X \j^{-1}$. Indeed, the expression~(\ref{eq:A12})
cannot generally be put into this form for any $W$. For
suppose that there exists some $W$
such that~(\ref{eq:A12}) and~(\ref{eq:A12fm}) are equivalent
for all self-adjoint $A^\prime$. Then
$R_W = \lambda\j R_Z R_Y^{-1} \j^{-1}$ for some $\lambda \in
{\Bbb C}$ which may be re-written as $\j^{-1}R_W(\j^{-1})^* =
\lambda R_Z R_Y$ using the unitarity of $\j R_Y$.
The LHS is self-adjoint, so the lemma in Appendix~A entails that
$ZZ^*$ and $YY^*$ must commute, which is a non-trivial condition on
$X$ and $Y$ when both are non-unitary. Hence in general, the
transformation~(\ref{eq:A12}) is not of the form~(\ref{eq:A12fm}).

Thus, for consistency to be maintained,
the transformation rule for observables between ${\cal H}^{\prime}$
and ${\cal H}^{\prime\prime}$ takes a different form from that which
holds between ${\cal H}$ and ${\cal H}^\prime$ or
${\cal H}^{\prime\prime}$. We regard this as an undesirable
feature of Anderson's proposal.

\section{The Unitary Dilation Proposal}
\label{sect:udp}

We begin by describing the theory of unitary
dilations~\cite{Nagy,Davis}. Let ${\cal H}_1,\ldots,{\cal H}_4$ be
Hilbert spaces and let $X$ be a bounded operator from ${\cal H}_1$ to
${\cal H}_2$. Then an operator $\hat{X}$ from ${\cal H}_1\oplus{\cal
H}_3$ to ${\cal H}_2\oplus{\cal H}_4$ is called a {\em dilation} of
$X$ if    $X=P_{{\cal H}_2}\hat{X}|_{{\cal H}_1}$ where $P_{{\cal
H}_2}$ is the orthogonal projector onto ${\cal H}_2$. In block matrix
form, $\hat{X}$ takes form
\begin{equation}
\hat{X} = \left(\begin{array}{cc} X & P \\ Q & R \end{array}\right).
\label{eq:dilfm}
\end{equation}
Our nomenclature follows that of Halmos \cite{Halmos}.

Given $X:{\cal H}_1\rightarrow{\cal H}_2$, one may ask whether $X$
possesses a {\em unitary} dilation. It turns out that such a
dilation always exists, although one must pass to indefinite inner
product spaces if the operator norm $\|X\|$ of $X$ exceeds unity.
One may construct a unitary dilation of $X$ as follows. Firstly, its
departure from unitarity may be quantified with the operators
$M_1=\openone-XX^*$ and $M_2=\openone-X^*X$. As a consequence of the
spectral theorem, we have the intertwining relations
\begin{equation}
X^*f(M_1)=f(M_2)X^*;\qquad X f(M_2)=f(M_1)X \label{eq:inter}
\end{equation}
for any continuous Borel function $f$. The closures of the ranges of
$M_1$ and $M_2$ are denoted ${\cal M}_1$ and ${\cal M}_2$
respectively.

For $i=1,2$, we now define ${\cal K}_i={\cal H}_i\oplus{\cal M}_i$,
equipped with the (possibly indefinite) inner product
$[\cdot,\cdot]_{{\cal K}_i}$ given by
\begin{equation}
\left[\left(\begin{array}{c} \varphi \\ \Phi\end{array}\right),
\left(\begin{array}{c} \psi \\
\Psi\end{array}\right)\right]_{{\cal K}_i}=
\langle\varphi\mid\psi\rangle +
\langle\Phi\mid {\rm sgn}\, M_i\Psi\rangle,
\end{equation}
where the inner products on the right are taken in ${\cal H}$ and
${\rm sgn}\, M_i=|M_i|^{-1}M_i$ where $|M_i|=(M_i^*M_i)^{1/2}$. It is
easy to show that ${\rm sgn}\, M_i$ is positive if $\|X\|\le 1$, in
which case $[\cdot,\cdot]_{{\cal K}_i}$ is positive definite;
however, for $\|X\|>1$, the inner products above are indefinite, and
${\cal K}_1$ and ${\cal K}_2$ are {\em Krein spaces} (for details on
the theory of operators in indefinite inner product spaces, see the
monographs \cite{Bognar,Azizov}). It is important to remember that
the ${\cal K}_i$ also have a positive definite inner product from
their original definition as a direct sum of Hilbert
spaces.\footnote{In fact, this inner product determines the topology
of ${\cal K}_i$.} Thus a bounded linear operator $A$ from ${\cal
K}_1$ to ${\cal K}_2$ has two adjoints: the Hilbert space adjoint
$A^*$, and the Krein space adjoint, which we denote $A^\dagger$. It
is a simple exercise to show that $A^\dagger$ is given by
\begin{equation}
A^\dagger = J_1 A^* J_2 ,
\label{eq:adj}
\end{equation}
where the operators $J_i$ defined on ${\cal K}_i$ are unitary
involutions given by $J_i=\openone_{{\cal H}_i}\oplus {\rm sgn}\,
(M_i)$.

Next, we define a dilation $\hat{X}:{\cal K}_1\rightarrow{\cal K}_2$
of $X$ by
\begin{equation}
\hat{X} =
\left(\begin{array}{cc} X & -{\rm sgn}\,(M_1)|M_1|^{1/2} \\
                                  |M_2|^{1/2} & X^*|_{{\cal M}_1}
                \end{array} \right),
\label{eq:dil}
\end{equation}
which has adjoint $\hat{X}^\dagger$ given by~(\ref{eq:adj}) as
\begin{equation}
\hat{X}^\dagger = \left(\begin{array}{cc}
X^* & {\rm sgn}\,(M_2)|M_2|^{1/2} \\ -|M_1|^{1/2} &
{\rm sgn}\,(M_1)X|_{{\cal M}_2}{\rm sgn}\,(M_2) \end{array}\right).
\end{equation}
It is then a matter of computation using the intertwining relations
to show that $\hat{X}^\dagger\hat{X}=\openone_{{\cal K}_1}$ and
$\hat{X}\hat{X}^\dagger=\openone_{{\cal K}_2}$. $\hat{X}$ is
therefore a unitary dilation of $X$.

The construction we have given is not unique. For suppose that ${\cal
N}_1$ and ${\cal N}_2$ are Krein spaces, and that $U_i:{\cal
M}_i\rightarrow{\cal N}_i$ are unitary (with respect to the
indefinite inner products). Then
\begin{equation}
\widetilde{X}=
\left(\begin{array}{cc} \openone & 0 \\ 0 & U_2 \end{array}\right)
\hat{X}
\left(\begin{array}{cc} \openone & 0 \\ 0 & U_1^\dagger
\end{array}\right)
\label{eq:free}
\end{equation}
is also a unitary dilation of $X$, mapping between ${\cal
H}\oplus{\cal N}_1$ and ${\cal H}\oplus{\cal N}_2$. Because this just
amounts to a redefinition of the auxiliary spaces, it carries no
additional physical significance. One may show that all other unitary
dilations of $X$ require the addition of larger auxiliary spaces than
the ${\cal M}_i$ (for example, one could dilate $\hat{X}$ further).
Thus $\hat{X}$ is the  minimal unitary dilation of $X$ up to
unitary equivalence of the above form.

Having described the general theory, let us now apply it to the case
of interest. For simplicity, we assume that the Hilbert spaces of
initial and final states are identical, so ${\cal H}_1={\cal
H}_2={\cal H}$.  We also assume that the evolution operator $X$ is
bounded with bounded inverse. If the initial hypersurface contains
regions which are causally separate from the CTC region, we assume
that $X$ has been normalised to be unitary on states localised in
such regions. We point out that such exterior regions may not exist
-- even if the CTC region is itself compact. Consider, for example, a
spacetime that is asymptotically (the universal cover of)
anti-de~Sitter space. In such a spacetime, hypersurfaces sufficiently
far to the future and far to the past of the CTC region will be
entirely contained within the CTC region's light cone and there will
be no exterior region on which to set up our normalisation. We may
normalise the evolution operator on hypersurfaces for which an
exterior region may be identified and extend arbitrarily to those
surfaces where no such region exists. Indeed, it is entirely possible
that every point in spacetime is contained in the light cone of the
CTC region; in this case we give up any attempt to find a `physical'
normalisation for the evolution operator.

The spaces ${\cal M}_1$ and ${\cal M}_2$ are defined as above. Note
that we have the polar decomposition $X=(XX^*)^{1/2}U$, where $U$ is
a {\em unitary} operator because $X$ is invertible. As a consequence
of the intertwining relations, we have
\begin{equation}
U M_2 = M_1 U
\end{equation}
and hence that ${\cal M}_1=U{\cal M}_2$. Thus the ${\cal M}_i$ are
isomorphic as Hilbert spaces. Moreover, $U$ is also unitary with
respect to the indefinite inner products on the auxiliary spaces
${\cal M}_1$ and ${\cal M}_2$, which follows from the identity
$U{\rm sgn}\, (M_2)={\rm sgn}\, (M_1)U$. We can therefore use the
freedom provided by equation~(\ref{eq:free}) to arrange that the
same auxiliary space is used both before and after the evolution.

Our proposal is the following. Given a non-unitary evolution $X$,
there exists an (indefinite) auxiliary space ${\cal M}$
(isomorphic to the ${\cal M}_i$) and a unitary
dilation $\widetilde{X}:{\cal K}\rightarrow{\cal K}$ of
$X$, where ${\cal K}={\cal H}\oplus{\cal M}$.
We regard this as describing the full physics of the situation:
on ${\cal K}$, the evolution is unitary, whilst its restriction to
the original Hilbert space ${\cal H}$ yields the non-unitary operator
$X$. The auxiliary space ${\cal M}$ represents degrees of
freedom localised within the CTC region, not directly accessible to
experiments outside.\footnote{{\em Indirectly}, we can infer their
presence by analysing $X$.}

Observables are defined as follows. Given any self-adjoint
operator $A$ on ${\cal H}$, we define the corresponding observable on
${\cal K}$:
\begin{equation}
\widetilde{A} = \left(\begin{array}{cc} A & 0 \\ 0 & 0
\end{array}\right).
\end{equation}
The form of $\widetilde{A}$ is chosen to
prevent the internal degrees of freedom being probed from outside.

Let us point out that many features of this proposal can only be
determined in the context of a particular evolution $X$ and therefore
a particular CTC spacetime. There are, however, various model
independent features of our proposal, which we discuss below.

{\noindent \bf  Predictability} Because the initial state involves
degrees of freedom not present on the initial hypersurface (i.e., the
component of the wavefunction in ${\cal M}$), it is clear that -- as
far as physical measurements are concerned -- there is some loss of
predictability in the final state. This problem can be circumvented
by the requirement that the initial state should have no component in
${\cal M}$.

{\noindent \bf Expectation Values} Let us examine the evolution of
the expectation value of $\widetilde{A}$. On the premise that the
initial state has no component in ${\cal M}$ and takes the vector
form $(\psi,0)^T$, the initial expectation value of $\widetilde{A}$ is
\begin{equation}
\frac{\left[\left(\begin{array}{c} \psi \\
0\end{array}\right),
\widetilde{A}\left(\begin{array}{c} \psi \\ 0\end{array}\right)
\right]_{{\cal K}_2}}{\left[
\left(\begin{array}{c} \psi \\ 0\end{array}\right),
\left(\begin{array}{c} \psi \\ 0\end{array}\right)
\right]_{{\cal K}_2}}=
\frac{\langle \psi\mid A\psi\rangle}{\langle\psi\mid\psi\rangle},
\end{equation}
i.e., the expectation value of $A$ in state $\psi$. After evolution,
the expectation value is
\begin{equation}
\frac{\left[\widetilde{X}\left(\begin{array}{c} \psi \\
0\end{array}\right),
\widetilde{A}\widetilde{X}\left(\begin{array}{c} \psi \\
0\end{array}\right) \right]_{{\cal K}_2}}{\left[\widetilde{X}
\left(\begin{array}{c} \psi \\ 0\end{array}\right),
\widetilde{X}\left(\begin{array}{c} \psi \\ 0\end{array}\right)
\right]_{{\cal K}_2}}=
\frac{\langle X\psi\mid AX\psi\rangle}{\langle\psi\mid\psi\rangle}.
\end{equation}
It is important to note that both denominators
are equal to $\|\psi\|^2$ (because the full evolution is
unitary) -- this removes many of the problems encountered in
Section~\ref{sect:nuqm}.

In particular, let us return to the problem noted by Jacobson
\cite{Jacobson}, writing ${\cal R}$ for the region spacelike
separated from the CTC region, and taking $X$ to be the evolution
from states on $\sigma_-$ to states on $\sigma_+$. We assume (as in
\cite{Jacobson}) that $X$ acts as the identity on ${\cal H}_{\cal R}$,
the subspace of states supported in ${\cal R}$. Any local observable
associated with ${\cal R}$ should vanish on the orthogonal complement
of ${\cal H}_{\cal R}$ in ${\cal H}$: accordingly, it follows that
$X^*AX=A$, and hence that the expectation value  is independent of
the choice of hypersurface ($\sigma_+$ or $\sigma_-$) on which it is
computed. Thus Jacobson's ambiguity is avoided for all local
observables associated with regions spacelike separated from the
causality-violating region. More generally, it is avoided for all
observables $A$ such that $A=X^*AX$. This is satisfied if the range
of $A$ is contained in ${\cal U}=\ker M_1\cap\ker M_2\subset{\cal H}$
and $A$ commutes with the restriction $X|_{\cal U}$ of $X$ to ${\cal
U}$.

In addition, the breakdown of the Copenhagen interpretation noted in
\cite{FPS1} is avoided as a direct consequence of the unitarity of
$\widetilde{X}$.

{\noindent \bf  Time Reversal} Let us suppose the existence of an
anti-unitary involution $T$ on ${\cal H}$ implementing time reversal.
The {\em time reverse} $X_{\rm rev}$ of $X$ is given by $X_{\rm
rev}=TXT^{-1}$; $X$ is said to be {\em time reversible} if $X_{\rm
rev}=X^{-1}$. We would like to understand how the time reversal
properties of $\hat{X}$ are related to those of $X$. For convenience
we will work in terms of $\hat{X}$; the discussion may be rephrased
in terms of $\widetilde{X}$ by inserting suitable unitary operators
between the ${\cal M}_i$ and ${\cal M}$.

First, we must define the time reversal of $\hat{X}$. The natural
definition is
\begin{equation}
(\hat{X})_{\rm rev} =
\left(\begin{array}{cc} T & 0 \\ 0 & T|_{{\cal M}_2}
\end{array}\right)
\hat{X}
\left(\begin{array}{cc} T^{-1} & 0 \\ 0 & (T|_{{\cal M}_1})^{-1}
\end{array}\right),
\end{equation}
which entails that time reversal and dilation commute in the
sense that $(\hat{X})_{\rm rev} = \widehat{X_{\rm rev}}$. However,
because dilation and inversion do not commute (i.e., $(\hat{X})^{-1}
\not = \widehat{X^{-1}}$) unless $X$ is unitary, we find that
a time reversible evolution $X$ does not generally yield a
time reversible dilation:
\begin{equation}
(\hat{X})_{\rm rev} = \widehat{X_{\rm rev}} = \widehat{X^{-1}}
\not = (\hat{X})^{-1}.
\end{equation}
Thus if $X$ is non-unitary and time
reversible, then $\hat{X}$ is not time reversible. On the other hand,
suppose that
$\hat{X}$ is time reversible. Then $\widehat{X_{\rm
rev}}=\widehat{X^*}$ from which it follows that $X$ would obey the
modified reversal property $X_{\rm rev}=X^*$.
It would be interesting to determine, for concrete CTC models,
whether $X$ obeys the usual time reversal property
$X_{\rm rev}=X^{-1}$ or the modified property $X_{\rm rev}=X^*$ (of
course it might not obey either property).

To summarise this section, we have seen how unitarity can be
restored  using the method of unitary dilations, thereby removing the
problems associated with non-unitary evolutions. Any observable on
${\cal H}$ defines an observable in our proposal.

\section{Composition}
\label{sect:comp}

We have described how a single non-unitary evolution may be dilated
to a unitary evolution between enlarged inner product spaces.
In what sense does our proposal respect the composition of two (or
more) non-unitary evolutions?

Let us consider two evolutions $X$ and $Y$
on ${\cal H}$ and their composition $XY$.
We define the $M_i$ and
${\cal M}_i$ as before and introduce $N_1=\openone-YY^*$,
$N_2=\openone-Y^*Y$ and ${\cal N}_i=\overline{{\rm Ran}\, N_i}$ to be
the closure of the range of $N_i$ for $i=1,2$. As before, we can
construct dilations $\hat{X}$ and $\hat{Y}$. However, because
$\hat{X}: {\cal H}\oplus{\cal M}_1\rightarrow
{\cal H}\oplus{\cal M}_2$ and
$\hat{Y}: {\cal H}\oplus{\cal N}_1\rightarrow
{\cal H}\oplus{\cal N}_2$, it is not immediately apparent how the
dilations may be composed. The solution is to dilate both $\hat{X}$
and $\hat{Y}$ further, as follows:
$\check{Y}:{\cal H}\oplus{\cal M}_1\oplus{\cal N}_1\rightarrow
{\cal H}\oplus{\cal M}_1\oplus{\cal N}_2$ is given by
\begin{equation}
\check{Y}=\left(\begin{array}{ccc}
Y & 0 & -{\rm sgn}\, N_1 |N_1|^{1/2} \\
0 & \openone_{{\cal M}_1} & 0 \\
|N_2|^{1/2} & 0  & Y^*|_{{\cal N}_1}
\end{array}\right),
\end{equation}
and $\check{X}:{\cal H}\oplus{\cal M}_1\oplus{\cal N}_2\rightarrow
{\cal H}\oplus{\cal M}_2\oplus{\cal N}_2$ is given by
\begin{equation}
\check{X}=\left(\begin{array}{ccc}
X &  -{\rm sgn}\, M_1 |M_1|^{1/2} & 0 \\
|M_2|^{1/2} & X^*|_{{\cal M}_1} & 0\\
0 & 0 & \openone_{{\cal N}_2}
\end{array}\right) .
\end{equation}
\widetext
The product $\check{X}\check{Y}$ is given by
\begin{equation}
\check{X}\check{Y}=\left(\begin{array}{ccc}
XY & -{\rm sgn}\, M_1 |M_1|^{1/2} & -X{\rm sgn}\, N_1 |N_1|^{1/2}\\
|M_2|^{1/2}Y & X^*|_{{\cal M}_1}
& -|M_2|^{1/2}{\rm sgn}\, N_1 |N_1|^{1/2} \\
|N_2|^{1/2} & 0 & Y^*|_{{\cal N}_1}
\end{array}\right),
\end{equation}
and is a unitary dilation of $XY$, mapping from ${\cal H}\oplus{\cal
M}_1\oplus{\cal N}_1$ to ${\cal H}\oplus{\cal M}_2\oplus{\cal N}_2$.

This state of affairs is quite natural: we have argued that each CTC
region carries with it its own auxiliary space (isomorphic to the
${\cal M}_i$ and the ${\cal N}_i$); one would therefore expect that
the combined evolution should be associated with the direct sum of
these auxiliary spaces. However, in order to show how our proposal
respects composition, we need to show how the product
$\check{X}\check{Y}$ is related to the dilation $\widehat{XY}$
arising from the prescription~(\ref{eq:dil}). To this end, we
introduce
$P_1=\openone-XYY^*X^*$, $P_2=\openone-Y^*X^*XY$ and ${\cal
P}_i=\overline{{\rm Ran}\, P_i}$. Note that
\begin{equation}
P_1=M_1+XN_1X^* \quad  {\rm and}\quad P_2=N_2+Y^*M_2Y.
\label{eq:iden}
\end{equation}
Now let
\begin{equation}
Q_1=\left(\begin{array}{c} |M_1|^{1/2} \\ |N_1|^{1/2}X \end{array}
\right)
\quad {\rm and}\quad
Q_2=\left(\begin{array}{c} |M_2|^{1/2}Y \\ |N_2|^{1/2} \end{array}
\right),
\end{equation}
and define $U_i$ ($i=1,2$) on ${\rm Ran}\, |P_i|^{1/2}\subset{\cal
P}_i$ by $U_i=Q_i|P_i|^{-1/2}$. The $U_i$ are easily seen to be
isometries (with respect to the appropriate inner products) from
their domains into ${\cal M}_i\oplus{\cal N}_i$ such that
$Q_i|_{\overline{{\rm Ran}\, P_i}}=U_i|P_i|^{1/2}$. Provided that
${\cal Q}_i=\overline{Q_i\overline{{\rm Ran}\, P_i}}$ is
orthocomplemented in ${\cal M}_i\oplus{\cal N}_i$ (in the indefinite
inner product), one may then show that
\begin{equation}
{\rm P}_{{\cal H}\oplus
{\cal Q}_2}\check{X}\check{Y}|_{{\cal H}\oplus{\cal Q}_1}=
\left(\begin{array}{cc} \openone & 0 \\ 0 & U_2 \end{array}\right)
\left(\begin{array}{cc}
XY & -{\rm sgn}\, P_1 |P_1|^{1/2} \\
|P_2|^{1/2} & (XY)^*|_{{\cal P}_1}
\end{array}\right)
\left(\begin{array}{cc} \openone & 0 \\ 0 & U_1^\dagger
\end{array}\right),
\end{equation}
where ${\rm P}_{{\cal H}\oplus{\cal Q}_2}$ is the orthoprojector onto
${\cal H}\oplus{\cal Q}_2$. Thus $\check{X}\check{Y}$ is a dilation
of an operator isometrically equivalent to $\widehat{XY}$. The
isometries act non-trivially only on the auxiliary spaces and have no
physical significance. The extra dimensions introduced by the
dilation are also to be expected because the combined evolution
$Z=XY$ may be factorised in many different ways; hence the two
individual evolutions carry more information than their combination.
\narrowtext

The assumption that the ${\cal Q}_i$ are orthocomplemented is easily
verified if the operators $U_i$ are bounded, for in this case, they
may be extended to unitary operators on the whole of ${\cal P}_i$.
Then ${\cal Q}_i$ is the unitary image of a Krein space and is
orthocomplemented by Theorem~VI.3.8 in \cite{Bognar}. $U_1$ is bounded
if there exists $K$ such that $\|P_1\psi\|<\epsilon$ only if
$\|M_1\psi\|+\|N_1X\psi\|<K\epsilon$ for all sufficiently small
$\epsilon>0$. Similarly, $U_2$ is bounded if $\|P_1\psi\|<\epsilon$
only if $\|M_2Y\psi\|+\|N_2\psi\|<K\epsilon$ for all sufficiently
small $\epsilon>0$. Physically, this equates to the reasonable
condition that the combined evolution can be `almost unitary' on a
given state only if the individual evolutions are also `almost
unitary'.

As a particular instance of the above, we consider the case where $Y$
is unitary. The $N_i$ therefore vanish and the ${\cal N}_i$ are
trivial; in addition, $P_1=M_1$ and $P_2=Y^*M_2Y$. The operator
$\check{Y}$ is
\begin{equation}
\check{Y}=\left(\begin{array}{cc}
Y & 0 \\ 0 & \openone_{{\cal M}_1} \end{array}\right)
\end{equation}
and $\check{X}=\hat{X}$. The combined evolution is thus
\begin{equation}
\check{X}\check{Y} = \hat{X}\left(\begin{array}{cc}
Y & 0 \\ 0 & \openone_{{\cal M}_1} \end{array}\right)
\end{equation}
which is unitarily equivalent to $\widehat{XY}$ in the sense that
\begin{equation}
\hat{X}\left(\begin{array}{cc}
Y & 0 \\ 0 & \openone_{{\cal M}_1} \end{array}\right)=
\left(\begin{array}{cc}
\openone & 0 \\ 0 & Y \end{array}\right)\widehat{XY}.
\end{equation}
We emphasise that the first factor on the RHS has no physical
significance and is merely concerned with mapping the auxiliary
spaces ${\cal P}_2$ to ${\cal M}_2$ in a natural way.

To conclude this section, we make three comments. Firstly, note that
if $A$ belongs to the class of observables which avoid the Jacobson
ambiguity for each CTC region individually, then it also avoids this
ambiguity for the combined evolution; for if $A=X^*AX=Y^*AY$, then
certainly $A=Y^*X^*AXY$. Thus there is no `multiple Jacobson
ambiguity'. Secondly, in our proposal one does not need to know the
past history of the universe in order to evolve forward from any given
time, because the auxiliary degrees of freedom associated with one CTC
region are essentially passive `spectators' during the evolution
associated with any other such region. This is in contrast with the
composition rule proposed by Anderson~\cite{Anote}. Thirdly, one might
ask~\cite{Aques} what would happen if the non-unitary evolution was
continuous rather than occurring in discrete steps. This question
could be tackled using a suitable generalisation of the theory of
unitary dilations of semi-groups discussed by Davies~\cite{Davies}.

\section{Conclusion}
\label{sect:conc}

We have examined Anderson's proposal~\cite{Arley} for restoring
unitarity to quantum evolution in CTC spacetimes,
and noted an operational problem arising when one considers the
composition of two or more non-unitary evolutions. Instead, we have advocated a
new method for the restoration of unitarity, based on the
mathematical theory of unitary dilations, which does respect
composition under certain reasonable conditions. Because unitarity is
restored on the full inner product space, problems associated
with non-unitary evolutions such as Jacobson's ambiguity are avoided.

Our philosophy here has been to regard the non-unitarity of $X$ as a
signal that the full physics (and a unitary evolution) is being
played out on a larger state space than is observed. This bears some
resemblance to the situation in special relativity, where time
dilation signals that one must pass to spacetime (and an indefinite
metric) in order to restore an orthogonal transformation between
reference frames. (Indeed, the Lorentz boost in two dimensional
Minkowski space is precisely an orthogonal dilation of the time
dilation effect).

For our case of interest, the physical picture is that the
auxiliary space ${\cal M}$ corresponds to degrees of freedom within
the CTC region. Non-unitarity of the evolution signals that a particle
cannot pass through the CTC region unscathed: part of the initial
state becomes trapped in the auxiliary space corresponding to the
CTCs. A similar conclusion is espoused by three of the authors of
\cite{GPPTa}.

In the case in which $X$ has norm less than or equal to unity (so
that the full space ${\cal K}$ has a positive definite inner
product), this effect has a relatively simple interpretation. Namely,
there is a non-zero probability that an incident particle will never
emerge from the CTC region. To see how this can occur, we note that
computations of the propagator (see particularly \cite{Pol2}) proceed
by requiring consistency of the evolution round the CTCs. We suggest
that part of the incident state becomes trapped in order to
achieve this consistency.

On the other hand, perturbative calculations in $\lambda\phi^4$
theory by Boulware \cite{Boul} suggest that $\|X\|$ could well exceed
unity. In this case, ${\cal K}$ is an indefinite Krein space, and it
would apparently be possible that the `probability' of the particle
escaping from the CTC region could be greater than one.  In
principle, one might try to avoid this by seeking natural positive
definite subspaces of the initial and final Krein spaces. The obvious
choice would be to take the initial Hilbert space to be ${\cal H}$
and the final Hilbert space to be the image of ${\cal H}$ under
$\widetilde{X}$. However, this may lead to some problems in defining
observables on the final Hilbert space. If one decides to face the
problem directly (which seems preferable), one would be forced to
conclude that CTCs are incompatible with the twin requirements of
unitarity {\em and} a Hilbert space structure. The initial and final
state spaces would naturally be Krein spaces. This would not be
entirely unexpected: studies of quantum mechanics on the `spinning
cone' spacetime \cite{JdSG} have concluded that the inner product
becomes indefinite precisely inside the region of CTCs.
`Probabilities' greater than unity would denote the breakdown of the
theory in a manner analogous to the Klein paradox (see the extensive
discussion in the monograph of Fulling \cite{Full}) in which strong
electrostatic fields force the Klein-Gordon inner product to be
indefinite. In our case, it is the geometry of spacetime
which leads us to an indefinite inner product. We expect that
particle creation would occur in this case, as it does in the usual
Klein paradox.

The Klein paradox can be resolved by treating the electromagnetic
field as a dynamic field, rather than as a fixed external field.
Particles are created in a burst as the field collapses (unless it is
maintained by some external agency). In our case it seems reasonable
that, in the context of a full quantum theory of gravity, a burst of
particle creation occurs and the CTC region collapses. This is
essentially the content of Hawking's Chronology Protection Conjecture
\cite{Hawking}. Thus the emergence of Krein spaces in our proposal
may be interpreted as a signal for the instability of the CTC
spacetime.

Finally, our treatment has been entirely in terms of states and
operators; it would be interesting to see how it translates into
density matrices and the language of generalised quantum mechanics
\cite{GMH}.

\acknowledgments

We thank Lloyd Alty, Arlen Anderson, Mike Cassidy, Andrew Chamblin,
Atsushi Higuchi, Seth Rosenberg and John Whelan for valuable
discussions. We also thank Malcolm Perry for useful discussions
concerning ref. \cite{GPPTa}.  In addition, CJF thanks Churchill
College, Cambridge, the Royal Society and the Schweizerischer
Nationalfonds for financial support.

\appendix

\section{}

In this appendix, we prove the following

{\noindent \bf Theorem} {\em Suppose $X$ and $Y$ are bounded with
bounded inverses. Then $U_{XY}=\lambda U_X U_Y$ if and only if
$X^*X$ commutes with $YY^*$ and $\lambda =1$.}

{\noindent \em Proof:}
Starting with the sufficiency, we note that $Z=(X^*)^{-1}
(X^*X)^{1/2}(YY^*)^{-1/2}X^{-1}$ is positive and squares to give
$(XYY^*X^*)^{-1}$ (using the commutation property). It follows that
$Z$ is equal to the unique positive square root of $(XYY^*X^*)^{-1}$
and hence that
\begin{equation}
U_{XY} = (XYY^*X^*)^{-1/2}XY = (X^*)^{-1} (X^*X)^{1/2}
(YY^*)^{-1/2}Y.
\end{equation}
Using the fact that $(X^*)^{-1} (X^*X)^{1/2}=U_X$, we have proved
sufficiency.

To demonstrate necessity, we note that $U_{XY}=\lambda U_X U_Y$ only
if
\begin{equation}
X^*(XYY^*X^*)^{-1/2}X = \lambda (X^*X)^{1/2}(YY^*)^{-1/2}.
\label{eq:only}
\end{equation}
It follows that the RHS must be self-adjoint and positive. We now
apply the following Lemma:

{\noindent \bf Lemma} {\em Suppose that $A$ and $B$ are bounded
with bounded inverses and self-adjoint, and suppose that $\alpha AB$
is self-adjoint and positive for some $\alpha\in{\Bbb C}$,
$\alpha\not=0$. Then $\alpha=\pm 1$ and $A$ and $B$ commute.}

{\noindent\em Proof:} Because $\alpha AB$ is self-adjoint, we have
\begin{equation}
\alpha AB = \alpha^* BA.
\label{eq:comm}
\end{equation}
Now note that
\begin{eqnarray}
\alpha^*(\alpha AB- z)^{-1} &=&
\alpha^*(\alpha^* BA-z)^{-1} \nonumber \\
&=& \alpha B
(\alpha AB -z\alpha^*/\alpha)^{-1}B^{-1}.
\end{eqnarray}
Because $AB$ has non-empty spectrum on the positive real axis and
because the resolvent $(\alpha AB- z)^{-1}$ is an analytic
operator valued function of $z$ in ${\Bbb C}\backslash{\Bbb R}^+$,
we conclude  that $\alpha^*/\alpha$ must be real and positive.
Accordingly, $\alpha=\pm 1$ and equation~(\ref{eq:comm}) implies
that $A$ and $B$ commute. $\square$

In our case, this implies that $\lambda = \pm 1$ and that $X^*X$
commutes with $YY^*$. Moreover, because the two square roots on the
RHS of
equation~(\ref{eq:only}) are positive and commute, we conclude that
$\lambda = 1$ in order that the RHS be positive. $\square$

\section{}

Here, we consider another possible method for the restoration of
unitarity which, however, suffers from problems related to Jacobson's
ambiguity.
Instead of focussing on direct sums of Hilbert spaces, this proposal
uses tensor products and always maintains a positive definite inner
product. We start with $X:{\cal H}\rightarrow {\cal H}$, bounded with
bounded inverse and non-unitary as before, and define a new Hilbert
space ${\cal H}_X=(\openone\otimes X)\Sigma$, where
$\Sigma\subset{\cal H}\otimes{\cal H}$ is the closure of the space of
finite linear combinations of terms of form $\psi\otimes\psi$ for
$\psi\in{\cal H}$. Similarly, we define ${\cal
H}_{X^{-1}}=(\openone\otimes X^{-1})\Sigma$. Now define the operator
$\tilde{X}=X\otimes X^{-1}$ restricted to ${\cal H}_X$. Clearly,
$\tilde{X}(\psi\otimes X\psi)= \varphi\otimes X^{-1}\varphi$ where
$\varphi=X\psi$, and so $\tilde{X}:{\cal H}_X\rightarrow{\cal
H}_{X^{-1}}$. Moreover,
\begin{eqnarray}
\langle\tilde{X}(\psi\otimes X\psi)\mid\tilde{X}(\varphi\otimes
X\varphi)\rangle &=& \langle X\psi\otimes\psi\mid
X\varphi\otimes\varphi\rangle\nonumber\\ &=& \langle X\psi\mid
X\varphi\rangle \langle\psi\mid\varphi\rangle \nonumber \\
&=& \langle\psi\otimes X\psi\mid \varphi\otimes X\varphi\rangle
\end{eqnarray}
and therefore $\tilde{X}$ is a unitary operator from ${\cal H}_X$ to
${\cal H}_{X^{-1}}$.

Let us examine the structure of this proposal in more detail. First,
there is a natural transposition operation ${\cal T}$ on ${\cal
H}\otimes{\cal H}$: ${\cal T} (\varphi\otimes\psi)
=\psi\otimes\varphi$. It is easy to see that $\tilde{X}$ is the
restriction of ${\cal T}$ to ${\cal H}_X$: hence all the information
about $X$ is encoded into the definition of ${\cal H}_X$. Have we
lost any information in this process? Suppose ${\cal H}_X={\cal H}_Y$
for two distinct operators $X$ and $Y$. Then $\openone\otimes Z$ is a
bounded invertible linear map (though not necessarily unitary) of
$\Sigma$ onto itself, where $Z=X^{-1}Y$.  Because ${\cal T}$
restricts to the identity on $\Sigma$, we require $\psi\otimes Z\psi
= Z\psi\otimes\psi$ for each $\psi\in{\cal H}$. Taking an inner
product with $\phi\otimes\psi$ for some $\phi$, we obtain
\begin{equation}
\langle\phi\mid\psi\rangle\langle \psi\mid
Z\psi\rangle= \langle\phi\mid Z\psi\rangle
\langle\psi\mid\psi\rangle.
\end{equation}
Because $\phi$ is arbitrary, $\psi$ is therefore an eigenvector of
$Z$. But $\psi$ was also arbitrary and therefore $Z=\lambda\openone$
for some constant $\lambda\in{\Bbb C}\backslash\{0\}$. Thus $Y=\lambda
X$, so this construction loses exactly one scalar degree of freedom.
Effectively, we have lost the (scalar) operator norm $\|X\|$ of $X$,
but no other information.

We have therefore restored unitarity at the price of introducing
a second Hilbert space and correlations between the two.
The evolution on the large space is unitary. This fits well with the
picture of acausal interaction between the initial space and the CTC
region in its future. The physical interpretation is as follows: the
`time machine' contains a copy of the external universe, which evolves
backwards in time, starting with the final state of the quantum
fields and ending with their initial state. It is impossible to
prepare the initial state of the CTC region independently from the
initial state of the exterior quantum fields.

However, problems arise when observables are defined. Here,
observables on the initial space are naturally defined to be
self-adjoint operators on ${\cal H}\otimes{\cal H}$ with ${\cal H}_X$
as an invariant subspace (observables on the final space would have
${\cal H}_{X^{-1}}$ invariant). An operator of form $A\otimes B$ maps
${\cal H}_X$ to itself only if $B=XAX^{-1}$; combining this with the
requirement of self-adjointness, one finds that $A$ must commute with
$X^*X$ and its powers. Thus this proposal places restrictions on the
class of allowed observables.

The requirement that ${\cal H}_X$ be an invariant subspace for all
observables was adopted so that our space of initial states is
invariant under the unitary groups generated by observables (e.g.
translations). If we relax this, and define observables to be
self-adjoint operators on ${\cal H}\otimes{\cal H}$, it appears that
$A\otimes\openone$ corresponds naturally to the operator $A$ on
${\cal H}$. However, this suffers from the ambiguity pointed out by
Jacobson \cite{Jacobson}.

\end{document}